\documentclass{resonance}

\usepackage{amsfonts}
\usepackage{amsmath}
\usepackage{amssymb}
\newcommand{\bed}{\begin{displaymath}}
\newcommand{\eed}{\end{displaymath}}
\newcommand{\bei}{\begin{itemize}}
\newcommand{\eei}{\end{itemize}}
\newcommand{\bef}{\begin{figure}}
\newcommand{\eef}{\end{figure}}
\newcommand{\ben}{\begin{enumerate}}
\newcommand{\een}{\end{enumerate}}
\newcommand{\beq}{\begin{equation}}
\newcommand{\eeq}{\end{equation}}
\newcommand{\ber}{\begin{eqnarray}}
\newcommand{\eer}{\end{eqnarray}}

\newcommand{\msun}{\mbox{{\rm M}$_{\odot}$}}
\newcommand{\rsun}{\mbox{{\rm R}$_{\odot}$}}

\newcounter{attnctr} 

\begin{document}


\title{Gravity Defied \\
     From potato asteroids to magnetised neutron stars}
\secondTitle{II. The failed stars}
\author{Sushan Konar}

\maketitle
\authorIntro{\includegraphics[width=2cm,angle=90]{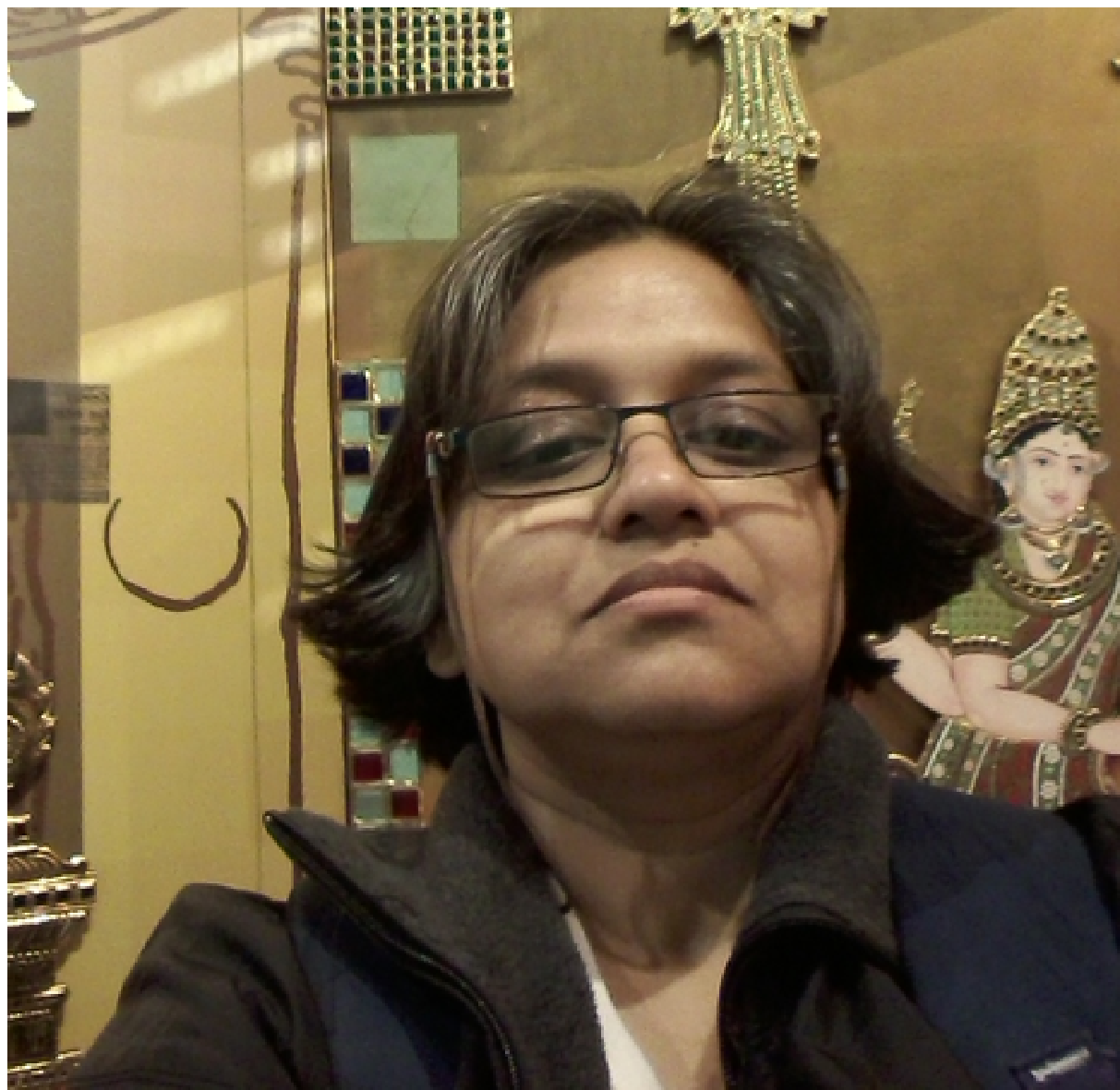}\\
Sushan Konar  works at  NCRA-TIFR, Pune. She  tries to  understand the
physics of stellar  compact objects (white dwarfs,  neutron stars) for
her livelihood and writes a blog about life in academia as a hobby.}
\begin{abstract}
  Gravitation, the  universal attractive  force, acts upon  all matter
  (and radiation)  relentlessly. Stable extended structures  can exist
  only  when gravity  is held  off by  other forces  of nature.   This
  series of articles explores this  interplay, looking at objects that
  just missed being stars in this particular installment.
\end{abstract}
\monthyear{January 2017}
\artNature{GENERAL  ARTICLE}

\section{Self-Gravitating Fluids}

\begin{figure}
  \caption{Balance of pressure in a self-gravitating fluid object.}
  \label{f_pressure}
%
\centering\includegraphics[width=4.0cm]{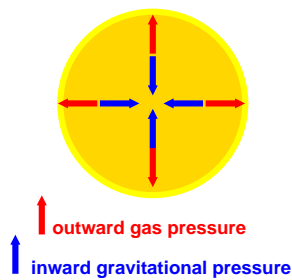}
\end{figure}

The first  resistance that nature  puts up against gravitation  is the
rigidity  of  solids.   We  have  seen how  small  rocky  objects  and
terrestrial planets  take recourse  to this  to retain  their extended
structures against gravitational  pull~\cite{konar17a}.  The situation
is entirely different in largely  fluid (gas/liquid) Jovian planets. A
fluid does not  have rigidity.  Instead, an element of  fluid would be
in  hydrostatic equilibrium  (at rest  or at  constant velocity)  when
external forces such  as gravity are balanced by  a pressure gradient.
In  other  words, in  a  self-gravitating  object, gravity  pulling  the
material inwards is balanced by the  internal pressure (due to heat or
quantum   effects)   differential   pushing  the   material   outwards
(Fig.~\ref{f_pressure}).


\subsection{Hydrostatic Equilibrium}
 
\begin{figure}
  \caption{Equilibrium of  a fluid  element inside a  self gravitating
    object, defining the equation of hydrostatic balance of pressure.}
  \label{f_hydro}
%
\centering\includegraphics[width=6.5cm]{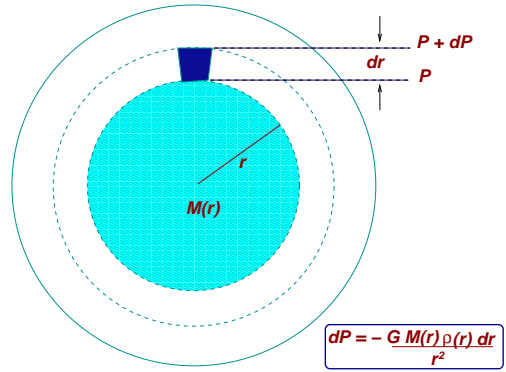}
\end{figure}

Consider the  fluid element  in a cylindrical  region of  length $dr$,
area $dA$,  at a distance  $r$ from  the centre of  a self-gravitating
object having a  density $\rho(r)$.  Then, the volume  of the cylinder
is $dr \, dA$  and its mass is $dm = \rho(r) \,  dr \, dA$.  The force
of gravity acting on the cylinder is,
\beq
F_{\rm G} = - \frac{G M(r) \, dm}{r^2} 
         = - \frac{G M(r) \, \rho(r) \, dr \, dA}{r^2} \,,
\eeq
where $M(r)$ is  the total mass contained within the  radius $r$. This
gravitational force  is balanced  by the  net difference  in pressure,
$dP$, from  above and below (Fig.~\ref{f_hydro}).   In equilibrium the
effective pressure force  should be equal to  the gravitational force,
giving us
\ber
dP = - \frac{G M(r) \, \rho(r) \, dr}{r^2}  \, \, \Rightarrow \, \,
\frac{dP}{dr} = - \frac{G M(r) \, \rho(r)}{r^2}  \,.
\eer
This equation of hydrostatic equilibrium  governs the structure of all
self-gravitating fluid  bodies, including  those of most  stars.  This
simple  form is, of course,  modified when  relativistic effects  become
important which we shall consider later.

%

\subsection{Gas Pressure}

\subsubsection{Kelvin-Helmholtz Mechanism}

The Kelvin-Helmholtz mechanism can be  thought of as a dynamic process
through  which a  self-gravitating body  attains successive  phases of
hydrostatic equilibrium.  To  begin with, a gas  cloud contracts under
the mutual gravitational attraction of its constituent particles. If a
spherical gas cloud of mass $M$  and initial radius $R_1$ contracts to
a final radius of $R_2$ then the  net change $\Delta E_{\rm G}$ in its
gravitational binding energy is,
\beq
\Delta E_{\rm G} = - G \, M \left(\frac{1}{R_1} - \frac{1}{R_2}\right)
\,,
\eeq
where $G$  is the  universal gravitational  constant. Evidently,  as a
result of contraction an amount of  energy is released. This goes into
increasing the random velocities  of the constituent particles raising
the average  temperature of the  object. (This is  why gravitationally
bound  objects are  known  to have  `negative  heat capacities'.)   The
increased temperature  increases the gas  pressure and the  system can
find  a  configuration  where  the inward  gravitational  pressure  is
balanced  by the  outward gas  pressure, i.e.  a state  of hydrostatic
equilibrium.

However, the  increased temperature  gives rise to  enhanced radiative
luminosity ($L \propto  T^4$, $L$ - surface luminosity,  $T$ - surface
temperature) from  the surface  which is a  process through  which the
object   starts   to  lose   the   heat   gained  from   gravitational
contraction. Evidently, this reduces the temperature and the resultant
gas pressure; eventually reaching a point when the gas pressure can no
longer support the  gravity.  This results in  an instability inducing
further contraction  of the  object and  a repeat  of the  whole cycle
itself.

[Kelvin and  Helmholtz proposed  this process  (late nineteenth century)
  to explain the energy source of the Sun. However, the total energy that
  can be  made available through  this process
  ($E_{\rm  G}^{\rm total} \sim G \msun^2/\rsun$)  would last only
  about  a few million years at  the present rate of  radiation from
  the Sun  whereas the fossil  records indicate  that the  Sun  has
  been  shining at  its present rate for a few billion  years at least.
  It was shown much later (1930s)  by Hans Bethe  that the  source of
  Sun's  energy is nuclear but that is a different story altogether.]

\subsubsection{Degeneracy Pressure}

As  a self-gravitating  object  continues through  phases of  repeated
contractions its density  keeps rising. When the density  is such that
the inter-particle spacing  ($ \propto n^{-1/3}$, $n$  - number density
of particles) becomes comparable to the thermal de Broglie wavelength,
given by
\beq
\lambda_{\rm de Broglie} \simeq \frac{h}{\sqrt{2 \pi m k_B T}} \,,
\eeq
(where $m$  is mass of the  particles, $k_B$ is the  Boltzmann constant
and $h$ is the Planck's constant) then quantum effects start having an
increasingly important role in the behaviour of the fluid.  It is then
that a  new agent for  resisting the gravity  is found in  the form
of quantum effects~\mfnote{Onset   of   quantum    effect
  :   $n^{-1/3}   \simeq \frac{h}{\sqrt{2 \pi m k_B T}}$}.

\begin{figure}[!b]
  \caption{Fermion energy distribution near absolute zero of temperature.}
  \label{f_fermi}
\vspace{-0.5cm} 
\centering\includegraphics[width=6.0cm]{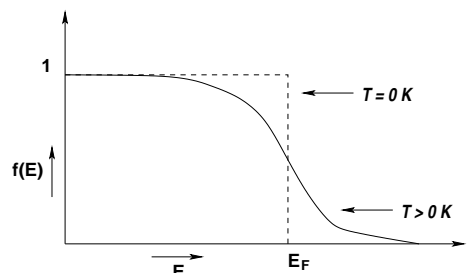}
\vspace{-0.5cm} 
\end{figure}

When  quantum  effects  become  important,  Fermions  (particles  with
half-integral spin following Fermi-Dirac  statistics) follow a special
energy  distribution given by,
\beq
f(E) = \frac{1}{e^{\frac{E - E_{\rm F}}{k_{\rm B}T}} + 1}
\eeq
where $E_{\rm  F}$ is known  as the  Fermi energy.  In  particular, at
absolute  zero of  temperature,  Fermions fill  up  all the  available
energy states  below $E_{\rm F}$ with  one (and only one)  particle in
each whereas the energy levels  above $E_{\rm F}$ have zero occupation
in them (Fig.~\ref{f_fermi}). Such a zero temperature Fermionic system
is  commonly  known as  a  {\em  degenerate}  system.

The  Fermi  temperature,  defined  as $T_{\rm  F}  =  E_{\rm  F}/k_B$,
provides an easy marker for the  behaviour of a Fermionic system. When
the physical  temperature $T$ of the  system is much smaller  than the
Fermi temperature ($T  << T_{\rm F}$) the behaviour of  the system can
be approximated  to that  of a  system at (or  near) absolute  zero of
temperature.  Interestingly, quite a few astrophysical objects (Jovian
planets, brown  dwarfs, white dwarfs  and neutron stars)  fulfill this
criterion and are therefore can be treated as zero temperature Fermion
systems.

For a non-relativistic, zero temperature Fermion system, the density of
particles (with a specific momentum $p$) is given by, 
\ber
n(p) \, dp &=& \frac{2. 4 \pi}{h^3} \, p^2 \, dp, \; \; p \leq p_{\rm F} \\
n(p) \, dp &=& 0, \; \; p > p_{\rm F}  \,,
\eer
where $p_{\rm F}$ is the momentum corresponding to $E_{\rm F}$.  The pressure
exerted by such a system is then,
\beq
P = \frac{m}{3} \int_0^\infty n(v) \, v^2 \, dv \;
  = \int_0^{p_{\rm F}} \frac{p^2}{3m} n(p) \, dp 
  = \left(\frac{3}{8 \pi}\right)^{\frac{2}{3}} \frac{h^2}{5 m} N^\frac{5}{3} \,,
\eeq
where $v (= p/m)$ is the velocity. And $N$ is the total number density
of particles given by
\beq
N = \int_0^{p_{\rm F}} n(p) \, dp = \frac{8 \pi}{3 h^3} \, p_{\rm F}^3\,.
\eeq
This  rather important  result for  non-relativistic degenerate  Fermi
systems  has  a special  significance  for  self gravitating  objects.
Consider a self-gravitating object of mass $M$ and radius $R$ composed
of a degenerate  Fermi gas.  From simple dimensional  estimates we can
find that the  central gravitational pressure of such  an object would
be,
\beq
P_{\rm G} \simeq \frac{G M^2}{R^4}\,.
\eeq
If this is to be balanced by the degeneracy pressure of the Fermi gas
then it implies that, 
\beq
\frac{M^2}{R^4} \propto N^{5/3} \Rightarrow M \, R3 = \mbox{constant}\,,
\eeq
(because $N$, the number density, scales  as $M/R^3$) giving us one of
the very  important relations  of planetary and  stellar astrophysics.
[Readers are advised  to look up standard  Statistical Mechanics texts
  \cite{pathr96,pal08} for details of Fermion systems.]

\section{The Jovian Planets}

\begin{figure}[!t]
%
  \caption{Interior of Jupiter as indicated by numerical simulations
    based on various observational data. Image taken from
    {\tt http://www.lpi.usra.edu/}}
  \label{f_jupiter}
  \vspace{-0.5cm}
\centering\includegraphics[width=7.5cm]{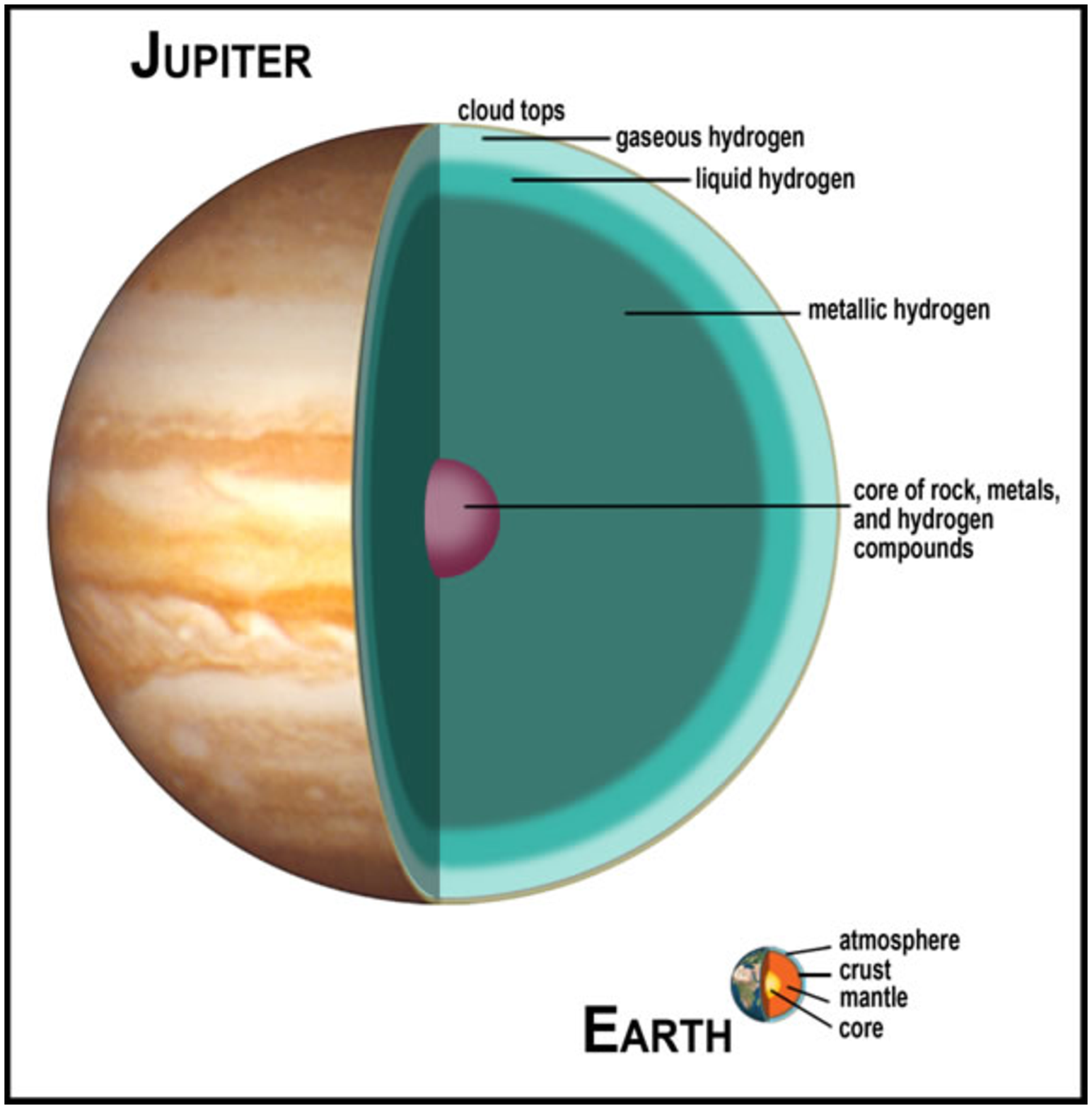}
\vspace{-0.5cm} 
\end{figure}

Jupiter, named  after the king of  ancient Roman gods, is  the largest
planet  of our  solar system  and  is thought  to consist  of a  dense
(rocky/icy?) core with  a mixture of elements, a  surrounding layer of
liquid  metallic  Hydrogen  with  some  Helium,  and  an  outer  layer
predominantly of molecular Hydrogen.  Evidently, the metallic Hydrogen
is  largest  component  of   this  planet  (Fig.~\ref{f_jupiter})  and
therefore plays the most significant role in the structure and stability
of Jupiter. Saturn, the next largest planet, is also expected to have
similar interior composition as Jupiter.

Both  Jupiter  and   Saturn  indicate  the  presence   of  a  positive
temperature gradient  from the  surface to the  interior. Conceivably,
the  higher  interior  temperatures  allow for  a  transition  of  the
molecular Hydrogen  (found in the  outer layers) to a  metallic phase.
This  metallic  state is  expected  to  be  quite similar  to  regular
terrestrial metals in which the  electrons form a degenerate Fermi gas.
It is  expected that  at the  end of  their Kelvin-Helmholtz  phase of
contraction  both  Jupiter  and  Saturn   would  settle  down  into  a
Hydrogen-degenerate phase  where the  resistance to  the gravitational
pressure would be supplied by the degeneracy pressure of the electrons
of the metallic Hydrogen (it needs to be noted that the pressure would
also have a small but non-zero contribution from the Hydrogen nuclei).

However, observations  indicate that  both Jupiter and  Saturn radiate
away more energy than they receive from the Sun. This clearly suggests
that both of  these planets are still in  their Kelvin-Helmholtz phase
of    contraction   and    heat    generation.   Interestingly,    the
Kelvin-Helmholtz process appears not to  account for the entire amount
of radiation from Saturn and suggests  that it may have another source
of power other than simple gravitational contraction.

\section*{Brown Dwarfs}

\begin{figure}
  \caption{Energy generation in the nuclear reaction as a result of mass
    defect. }
  \label{f_mdefect}
\vspace{-0.5cm} 
\centering\includegraphics[width=6.0cm]{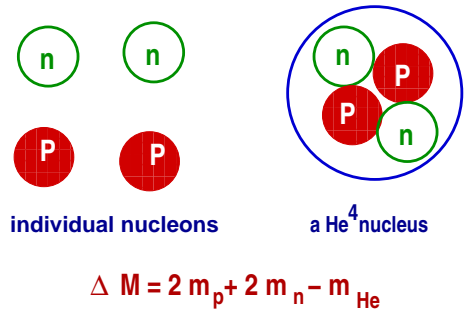}
\vspace{-0.5cm} 
\end{figure}

A star gets  its energy from nuclear burning,  combining four Hydrogen
(H) atoms  into one Helium (He$^4$)  atom, for most of  its life.  The
mass of  He$^4$ happens to  be slightly smaller  than the mass  of the
four  Hydrogen  atoms,  giving  rise  to a  small  {\em  mass  defect}
(meaning,  a deficit).  It  is this  defect that  shows  up as  energy
through the famous mass-energy equivalence  relation of Einstein ($E =
\Delta M \, c^2$) (Fig.~\ref{f_mdefect}).

\begin{figure}[!b]
%
  \caption{The $p-p$ chain, showing steps of conversion of Hydrogen into
    Helium. This is the process of energy generation active inside our Sun.}
  \label{f_fusion}
\vspace{-0.5cm} 
\centering\includegraphics[width=6.5cm]{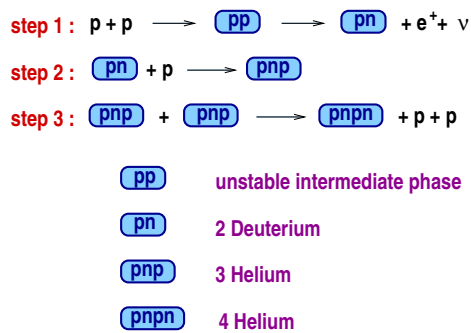}
\vspace{-0.5cm} 
\end{figure}
Now the  conversion of H into  He$^4$ actually happens in  a number of
steps, each requiring higher and  higher ambient temperature for it to
progress  (Fig.~\ref{f_fusion}).    When  a   self-gravitating  object
undergoes repeated  phases of Kelvin-Helmholtz contraction,  the final
temperature  reached depends  on its  mass. The  higher the  mass, the
higher  is the  final  temperature. Though  quite  heavy by  planetary
standards, the Jovian planets fall  far short of the magic temperature
(about a million degree Kelvin) to begin Hydrogen fusion.  Between the
Jupiters and the stars there exist another class of objects, the brown
dwarfs, which are  really {\em failed stars}. They  do achieve nuclear
fusion but  not of Hydrogen. Instead,  brown dwarfs are known  to burn
Deuterium (D$^2$, a heavier isotope of Hydrogen). 

In Deuterium fusion, a Deuterium nucleus  and a proton combine to form
a Helium-3  (He$^3$) nucleus. This occurs  at the second stage  of the
$p–p$  chain  of  the  Hydrogen  fusion, but  can  also  proceed  from
primordial Deuterium.   This onset  of Deuterium fusion  separates the
class  of brown  dwarfs from  that of  the Jovian  planets giving  the
maximum  mass of  gas  giants to  be about  0.012~\msun~(or  about 13
Jupiter masses). Whereas, the class  of brown dwarfs is separated from
the class of stars by the threshold for Hydrogen fusion at 0.08~\msun.
Once the Deuterium  burning gets over a brown dwarf  continues to cool
and ultimately the pressure support comes from the electron degeneracy
pressure of H and He$^3$.

%
\begin{figure}
  \caption{A schematic comparison of various stellar and sub-stellar
    objects. Picture taken from {\tt https://spaceplace.nasa.gov/}}
  \label{f_bd}
\vspace{-0.5cm} 
\centering\includegraphics[width=10.0cm]{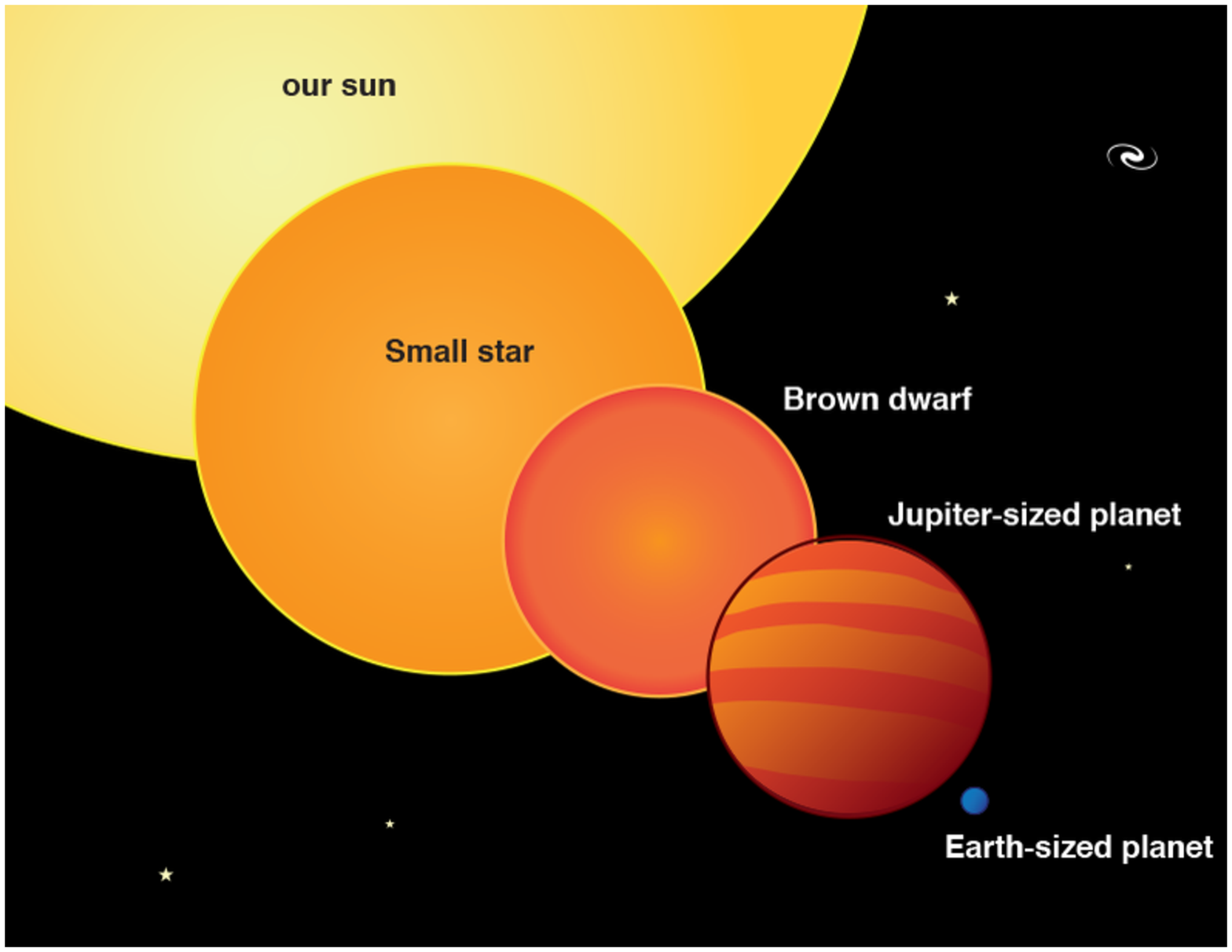}
\vspace{-0.5cm} 
\end{figure}

A  particularly  important  characteristic  of  brown  dwarfs  is  the
presence  of primordial  Lithium  in them.  Stars, achieving  Hydrogen
fusion temperature rapidly  deplete their Lithium. On  the other hand,
brown dwarfs (even massive brown  dwarfs) never reach this temperature
and therefore mostly retain their primordial Lithium.  The presence of
Lithium is therefore  used to distinguish candidate  brown dwarfs.

\section*{Acknowledgment}

My  thanks  to   G.   Srinivasan  for  setting  me  on   the  path  of
Fermi-degenerate  objects  by explaining  that  the  study of  stellar
evolution  actually begins  with Jupiter  (among other  such wonderful
concepts that  we'd encounter in the next installments of  this series)
and to  D.  Narasimha for introducing me to  the physics of  stars one
exciting summer thirty years ago. I would also like to express my deep
gratitude to the Resonance team for their patience and perseverance in
getting this article published after I miserably failed all possible
deadlines.

\end{document}